%% file: main.tex
\documentclass[twocolumn]{aastex63}
\turnoffeditone
\turnoffedittwo
\shorttitle{Environment of Quasar ULAS J1342+0928}
\shortauthors{Rojas-Ruiz et al.}

\usepackage{float}
\usepackage{natbib}
\usepackage{hyperref}
\usepackage[nointegrals]{wasysym}
\usepackage{ragged2e}
\usepackage{graphicx}
\usepackage{units}
\usepackage{amsmath}
\input{definitions}

\begin{document}

\title{Exploring the Mpc Environment of the Quasar ULAS J1342+0928 at $z = 7.54$}
\correspondingauthor{Sof\'ia Rojas-Ruiz}
\email{rojas@astro.ucla.edu}

\author[0000-0003-2349-9310]{Sof\'ia Rojas-Ruiz}\altaffiliation{Fellow of the International Max Planck Research School for Astronomy and Cosmic Physics at the University of Heidelberg (IMPRS--HD)}
\affiliation{Max-Planck-Institut f\"{u}r Astronomie, K\"{o}nigstuhl 17, D-69117, Heidelberg, Germany}\affiliation{Department of Physics and Astronomy, University of California, Los Angeles, 430 Portola Plaza, Los Angeles, CA 90095, USA}

\author[0000-0002-5941-5214]{Chiara Mazzucchelli}
\affiliation{Instituto de Estudios Astrof\'{\i}sicos, Facultad de Ingenier\'{\i}a y Ciencias, Universidad Diego Portales, Avenida Ejercito Libertador 441, Santiago, Chile.}

\author[0000-0001-8519-1130]{Steven L. Finkelstein}
\affiliation{The University of Texas at Austin, Department of Astronomy, \\
 2515 Speedway Boulevard Stop C-4636400, \\
Austin, TX 78712, USA}

\author[0000-0002-2931-7824]{Eduardo Ba\~nados}
\affiliation{Max-Planck-Institut f\"{u}r Astronomie, K\"{o}nigstuhl 17, D-69117, Heidelberg, Germany}

\author[0000-0002-6822-2254]{Emanuele Paolo Farina}
\affiliation{Gemini Observatory, NSF’s NOIRLab, 670 N A’ohoku Place, Hilo, Hawai'i 96720, USA}

\author[0000-0001-9024-8322]{Bram P.\ Venemans}
\affiliation{Leiden Observatory, Leiden University, PO Box 9513, 2300 RA, Leiden, The Netherlands}

\author[0000-0002-2662-8803]{Roberto Decarli}
\affiliation{INAF -- Osservatorio di Astrofisica e Scienza dello Spazio di Bologna, via Gobetti 93/3, I-40129, Bologna, Italy.}

\author[0000-0002-4201-7367]{Chris J. Willott}
\affiliation{NRC Herzberg Astronomy and Astrophysics, 5071 West Saanich Rd, Victoria, BC, V9E 2E7, Canada}

\author[0000-0002-7633-431X]{Feige Wang}
\affiliation{Steward Observatory, University of Arizona, 933 N Cherry Avenue, Tucson, AZ 85721, USA}

\author[0000-0003-4793-7880]{Fabian Walter}
\affiliation{Max-Planck-Institut f\"{u}r Astronomie, K\"{o}nigstuhl 17, D-69117, Heidelberg, Germany}

\author{Enrico Congiu} 
\affiliation{European Southern Observatory (ESO), Alonso de C\'ordova 3107, Casilla 19, Santiago 19001, Chile.}

\author[0000-0003-2680-005X]{Gabriel Brammer} 
\affiliation{Cosmic Dawn Center (DAWN), Denmark}\affiliation{Niels Bohr Institute, University of Copenhagen, Jagtvej 128, DK-2200 Copenhagen N, Denmark}

\author[0000-0002-6091-7924]{Peter Zeidler}
\affiliation{AURA for the European Space Agency (ESA), Space Telescope Science Institute, 3700 San Martin Drive, Baltimore, MD 21218, USA}


\begin{abstract}
Theoretical models predict that $z \gtrsim 6$ quasars are hosted in the most massive halos of the underlying dark matter distribution and thus would be immersed in protoclusters of galaxies. However, observations report inconclusive results. We investigate the 1.1 pMpc$^2$ environment of the $z = 7.54$ luminous quasar ULAS J1342+0928. We search for Lyman-break galaxy candidates (LBG) using deep imaging from \edit1{the {\it Hubble Space Telescope (HST)}}\ in the ACS/F814W, WFC3/F105W/F125W bands, and Spitzer/IRAC at 3.6 $\mu$m and 4.5 $\mu$m. We report a $z_{phot} = 7.69^{+0.33}_{-0.23}$ LBG with $mag_{F125W} = 26.41$ at $223$ projected-pkpc from the quasar. 
\replaced{Together with one dust-obscured galaxy previously found with ALMA at $27$ projected-pkpc and $z_\mathrm{[C\,{\sc II}]}=7.5341\pm0.0009$ \citep{venemans_kiloparsec-scale_2020}, suggest a galaxy overdensity around ULAS J1342+0928.}{\edit2{We find no \hst\ counterpart to one \cii-emitter }  previously found with ALMA at $27$ projected-pkpc and $z_\mathrm{[C\,{\sc II}]}=7.5341\pm0.0009$ \citep{venemans_kiloparsec-scale_2020}.}\
\edit1{We estimate the completeness of our LBG candidates using results from CANDELS/GOODS deep blank field searches sharing a similar filter setup. } We find that $>50\%$ of the $z\sim7.5$ LBGs with $mag_{F125W} >25.5$ are missed due to the absence of a filter redward of the Lyman-break in F105W, 
\replaced{that would produce more accurate UV colors of the candidates}{hindering the UV color accuracy of the candidates.}
\edit1{We conduct a } QSO-LBG clustering analysis revealing \edit1{a low LBG excess of $0.46^{+1.52}_{-0.08}$}  in this quasar field, \edit2{consistent with an average or low-density field. Consequently, }\edit1{this result does not present strong evidence of an LBG overdensity around ULAS J1342+0928. \edit2{Furthermore, we identify} two LBG candidates with a $z_{phot}$ matching a confirmed $z=6.84$ absorber along the line-of-sight to the quasar. } All these galaxy candidates are excellent targets for follow-up observations with \textit{JWST} and/or ALMA to confirm their redshift and physical properties.

\end{abstract}
\section{Introduction}\label{intro}

Understanding the formation of the first massive galaxies and black holes and their role in reionizing the universe is one of the main problems in modern cosmology. However, it is still challenging to identify these distant sources and subsequently characterize their properties. Quasars are the most luminous non-transient sources known and can be studied in detail at the earliest cosmic epochs \citep[e.g.][]{fan_quasars_2023}. Despite quasars being very rare sources ($\sim$ 1 per Gpc$^3$ at $t_{age} <$ 1 Gyr, \citealp{schindler_pan-starrs1_2023}), multiple observational efforts during the past decade have revealed a significant ($>400$) population of quasars in the epoch of reionization within the first billion years of the universe, at redshift $z>5.5$ \citep[e.g][]{venemans_discovery_2013, venemans_identification_2015,banados_pan-starrs1_2016,mazzucchelli_physical_2017,matsuoka_subaru_2019,reed_three_2019,yang_exploring_2019,gloudemans_discovery_2022,yang_desi_2023}. These observations evidence a dramatic decline of the spatial density of luminous quasars at $z > 6$ and suggest that we are closing into the epoch when the first generation of supermassive black holes (SMBHs) emerged
in the early universe \citep{wang_exploring_2019}.  

Only eight quasars are known at $z > 7$, and three are at $z>7.5$: 
J0313–1806 at $z = 7.64$ \citep{wang_luminous_2021}, J1342+0928 at $z = 7.54$ \citep{banados_800-million-solar-mass_2018}, and J1007+2115 at $z = 7.52$ \citep{yang_poniuaena_2020}. These early quasars are powered by $\gtrsim 10^8$ \Msun\ black holes \citep[e.g][]{yang_probing_2021,farina_x-shooteralma_2022} and the large majority reside in extremely
star-forming galaxies ($>$ 100 -- 1000 \Msun\,/yr; e.g. \citealt{venemans_kiloparsec-scale_2020}).
In order to sustain both the tremendous
black hole growth and the intense star formation, current theoretical models posit that these systems lie in highly
biased regions of the universe at that time, where gas can fragment and form a large number of surrounding
galaxies \citep[e.g.][]{springel_simulations_2005,volonteri_quasars_2006,costa_environment_2014}. These quasar environments could possibly host powerful sources of ionizing photons such as bright Lyman-$\alpha$ emitters, or have nearby halos hosting these galaxies \citep{overzier_lcdm_2009}. Consequently, these massive quasars are thought to be indicators of protoclusters defined as galaxy overdensities that will evolve by $z\sim0$ into the most massive ($\geq 10^{14}$ \Msun) virialized clusters \edit1{\citep{overzier_realm_2016}}. Studying the environment of quasars hosting SMBHs as early as at $z\sim7.5$ is crucial to understand the large-scale structure and the feeding of \added{ gas in }the first massive galaxies and black holes in the universe.

To probe the presence of such protoclusters, one can perform deep imaging observations to select galaxy candidates, and compare their number density to that observed in ``blank fields", i.e. field without a quasar. However, whether quasars at $z\sim6$ reside in overdense regions is heavily debated in the observational side of the literature. Discrepancies in these findings can be explained by the different observational techniques used to identify galaxies around quasars. \edit1{This includes photometric searches for} Lyman-break galaxies \citep[LBGs,][]{zheng_overdensity_2006,morselli_primordial_2014,simpson_no_2014,champagne_mixture_2023}, for Lyman-$\alpha$ emitters \citep[LAEs,][]{mazzucchelli_no_2017}, or for a combination of both \citep[e.g.,][]{ota_large-scale_2018}. \edit1{Also,} spectroscopic confirmations of galaxies \citep[e.g.,][]{bosman_three_2020,mignoli_web_2020}, or \cii\ emitters and sub-millimeter galaxy searches \citep[e.g.,][]{decarli_rapidly_2017,champagne_no_2018, meyer_constraining_2022} \edit1{have been undertaken in the literature. Recently, leveraging the capabilities of \jwst\ near-infrared spectra, a substantial influx of \oiii-emitting galaxies has been unveiled in the environments of $z\gtrsim5$ quasars \citep[][]{kashino_eiger_2023,wang_spectroscopic_2023}. Moreover, these studies encompass diverse physical areas and rely on different methods for evaluating the presence of an overdensity} \citep[e.g.,][]{overzier_conditions_2022}. Finally, the results are affected by cosmic variance given the handful of $z\sim6$ quasar fields inspected \citep{garcia-vergara_clustering_2019}.

The highest-redshift simulations available from \citet{costa_environment_2014} demonstrate that overdensities of (LBGs) and young LAEs around quasars up to $z\sim6.2$ can be probed within a 1.2 \edit1{proper-Mpc$^2$ (pMpc)$^2$} environment using the \hst\ ACS Wide Field Channel. The highest-redshift quasar whose environment has been studied so far, and using this observational strategy, is ULAS J1120+0641 at $z=7.1$ \citep{simpson_no_2014}. Given the rapidly decreasing number density of luminous quasars at $z>7$ \citep{wang_exploring_2019} where the formation of SMBHs posits challenges not only on theories of black hole formation but also on large-scale structure assembly \citep[e.g.,][]{habouzit_number_2016,habouzit_black_2016}, it is crucial to observationally inspect the environments of quasars at the highest-redshift known, i.e. $z \sim 7.5$. In this work, we search for LBG candidates at $z\sim7.5$ in the immediate $\sim1$ pMpc$^2$ environment of the $z=7.54$ quasar \Pisco, using deep imaging data collected with the {\it Hubble Space Telescope}\, (\hst), and \spitzer/Infrared Array Camera (IRAC). This quasar hosts one of the earliest and most massive SMBHs with a mass $\sim 0.9 \times 10^9$ \Msun, that is actively accreting at near Eddington rates with $L_{bol}/L_{Edd}\sim 1.1$ \citep{onoue_no_2020}. The host galaxy is already evolved with high amount of gas and dust resulting in SFR of $\sim150$\,\Msun\ yr$^{-1}$, with a metallicity comparable to the solar neighborhood \citep{novak_alma_2019}. Additionally, a study of the optical/NIR spectrum of \Pisco\ identified a strong absorber at $z=6.8$ on its line-of-sight \citep{simcoe_interstellar_2020}. This massive and active quasar in the early universe is an ideal candidate to now look for a galaxy overdensity and trace its large-scale structure.

\edit2{This paper is organized as follows: we describe the \hst\ data and their reduction in \S \ref{observations}, followed by the \hst\ photometry, noise calculation, and aperture corrections in \S \ref{photometry}. We also include available \spitzer/IRAC photometry (\S \ref{irac}).} The selection criterion and photometric-redshift analysis to create the final catalog of LBG candidates are described in \S \ref{selection}. Details on the properties of the resulting galaxy candidates are discussed in \S \ref{results}.The results, catalog completeness, and the interpretation of findings in relation to \replaced{galaxy overdensity}{the density of the quasar field} are discussed in \S \ref{discussion}. Finally, we summarize our results and provide further outlook in \S \ref{summary}. Throughout this article we adopt a cosmology with: H$_0$= 70 km~s$^{-1}$ Mpc$^{-1}$, $\Omega_M$= 0.3, $\Omega_\Lambda$ = 0.7. Using this cosmology, the age of the Universe is 679 Myr at the redshift of \Pisco, and $1"$ corresponds to 4.99 \edit1{proper-kpc (pkpc)}. All magnitudes provided are in the AB system.

\section{Observations}\label{observations}

Usually, at least three filters are occupied to identify galaxies in the epoch of reionization using the Lyman-break technique. The bluest filter serves to spot the spectral break in the galaxy continuum emission produced by the intergalactic medium (IGM) absorption. Hence, no or very little flux is expected to be detected in this filter. A contiguous filter is centered on the expected wavelength of the Lyman-break serving as the drop-out and detection band, and redder filters are used to observe the continuum emission. In this section, we describe the \hst\ data obtained to select LBG candidates in the environment of \Pisco\ and the reduction process.

\subsection{\hst\ data and reduction}
\begin{figure}
\centering
\includegraphics[width=\linewidth]{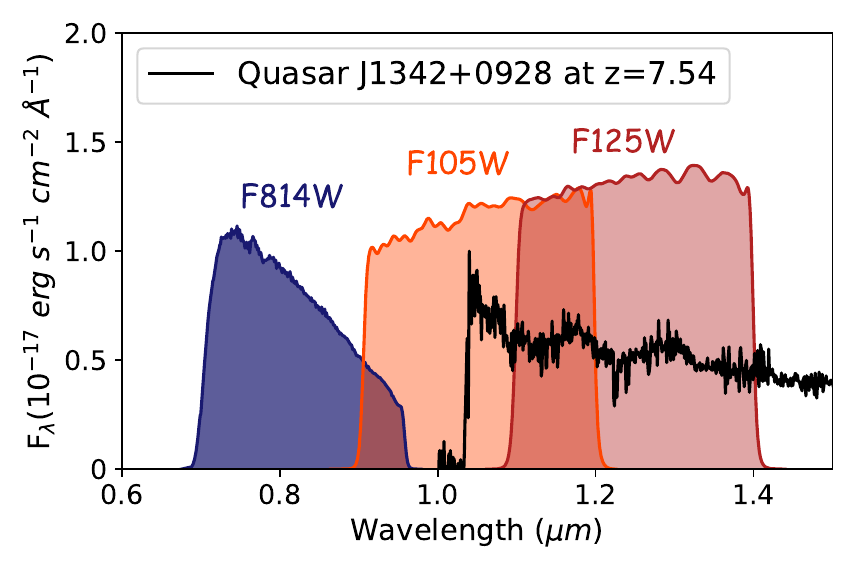}
\caption{The \hst\ filter coverage used to look for galaxy candidates in the field of \Pisco\ \edit2{is depicted, with} the spectrum of the quasar from \citet{banados_800-million-solar-mass_2018} overlaid. Galaxy candidates laying in close proximity to the quasar at $z\sim7.5$ \edit1{are expected to} be undetected in the WFC/F814W, begin to drop completely halfway through the WFC3/IR F105W filter, and be fully detected in the WFC3/F125W filter. 
}
\label{filters}
\end{figure}

We use observations obtained with the Advanced Camera for Survey (ACS) and Wide Field Camera 3 (WFC3) on board \hst\ between June 2018 and June 2019 (PI: Bañados, Prog ID:1165). We obtained data in the F814W (ACS, 13 orbits) serving as the non-detection filter, and the F105W and F125W filters (WFC3, 8 and 4 orbits each, respectively). To maximize the ACS surveyed area, WFC3 near-infrared imaging were observed in a $2\times2$ mosaic strategy. The final effective area covered to search for LBGs \edit1{ is computed based on the ACS/F814W image, as this area is covered by all three filters. The calculation masks out bad pixels in the weight map, resulting in an area of $12.28\, \mathrm{arcmin}^2$.} All filter transmission curves are presented in Figure \ref{filters}, with the rest-frame UV spectrum of \Pisco\ overlaid \citep{banados_800-million-solar-mass_2018}. The presented observations achieve a 5$\sigma$ limiting AB magnitudes of 28.20, 27.83, 27.46 in the F814W, F105W, and F125W bands, respectively,\edit1{ as calculated with a 0\farcs4-diameter circular aperture.}

\begin{figure*}[ht]
\centering
\includegraphics[width=\linewidth]{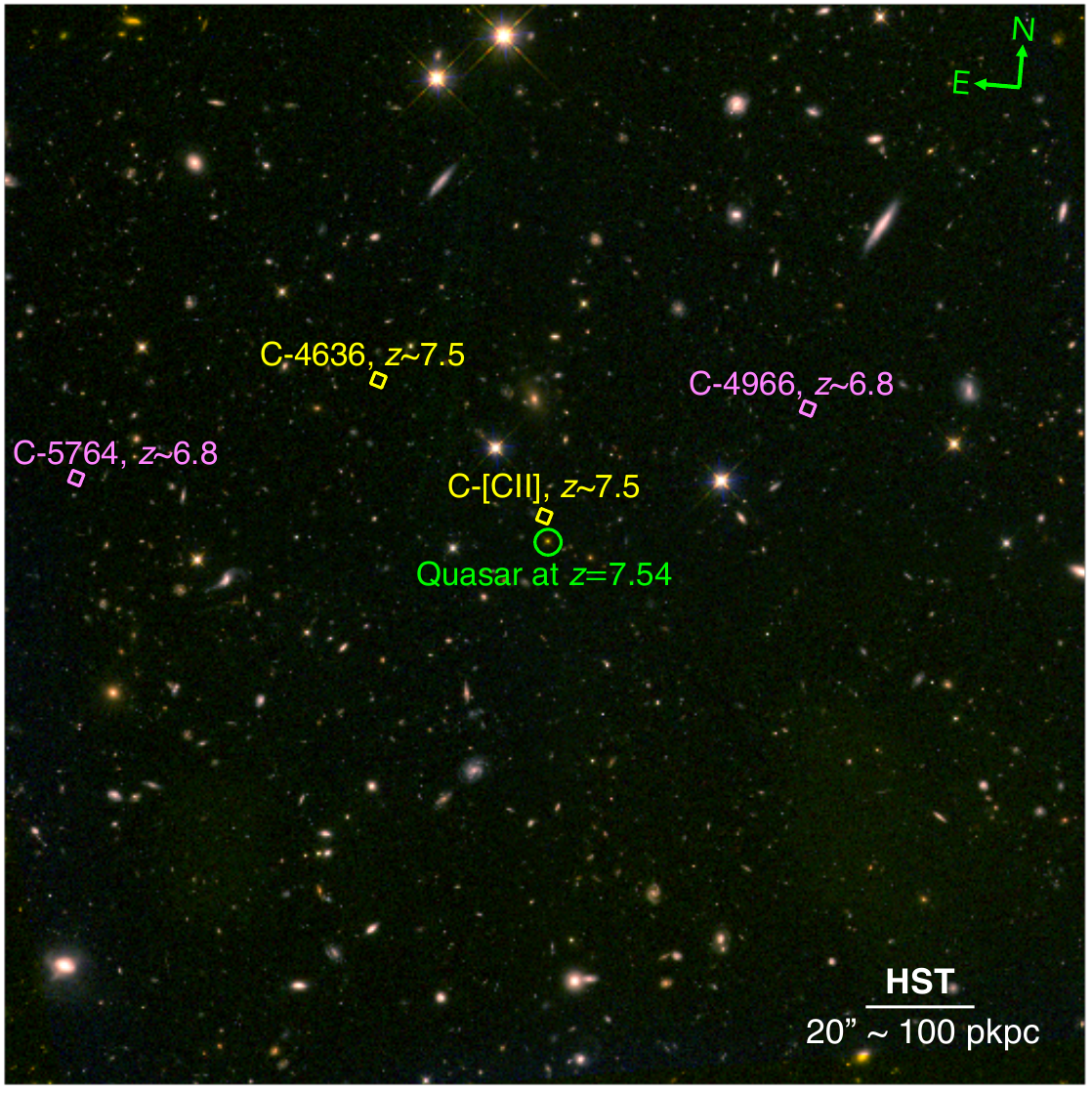}
\caption{\hst\ \textit{iYJ} RGB color-image of the field around the quasar \Pisco. The quasar is in the center as presented in the circled region. \edit2{Overlaid are the high-redshift galaxy candidates selected in this work as described in \S\ref{results}.} The source C-4636 \replaced{is an LBG candidate identified in this analysis to be at $z_{phot}=7.69$.}{is identified as an LBG candidate with a photometric redshift of $z_{\text{phot}}=7.69$.} Candidate C-\cii\ is a \cii\ emitter previously identified in \citet{venemans_kiloparsec-scale_2020} to be at $z_{\mathrm{[C\,{\sc II}]}}=7.5341\pm0.0009$. This candidate lacks \hst\ or \textit{Spitzer} counterpart emission, making it a dust-obscured candidate in the environment of the quasar. Additional LBG candidates in the observed field, C-4966 and C-5764 are at $z_{phot}=6.91$ and $6.89$, respectively. }
\label{rgb}
\end{figure*}

We use the bias subtracted, flat-fielded, and \replaced{cleaned for cosmic rays data-reduced images}{cosmic-ray cleaned, reduced images} provided by STScI, and implement an ad-hoc method to ensure a good astrometric match between the different filters \edit1{using} DrizzlePac.\footnote{\href{https://www.stsci.edu/scientific-community/software/drizzlepac.html}{https://www.stsci.edu/drizzlepac.html}} Indeed, such an alignment is nontrivial due to the small number of stars found in the field, which complicates standard reduction routines. We start by considering the  \textit{HST} WFC/F814W pipeline-reduced \textit{flc.fits} files, downloaded from the MAST archive\footnote{\url{https://archive.stsci.edu/}}. In order to create reference catalogs with enough sources, we run \texttt{Source Extractor} \citep{bertin_sextractor_1996} on each image, after cleaning them from cosmic rays contamination using the \textit{astrodrizzle} routine with \texttt{cosmic\_ray\_cr\_clean}=True. We use \textit{tweakreg} to align the uncleaned, original \textit{flc.fits} images, utilizing these \texttt{Source Extractor}-created reference catalogs, each containing $\sim$1000 sources. The final combined image in F814W is obtained using \textit{astrodrizzle}, with \texttt{skymethod}=`match' and \texttt{combine\_type}=`median'. We run again \texttt{Source Extractor} on the final, drizzled F814W image, and use this new catalog (4544 sources) as a reference to match the WFC3/F105W and F125W images to the F814W. In detail, we use \textit{tweakreg} on the F105W and F125W \textit{HST} pipeline-reduced \textit{flt.fits} files, with the F814W catalog as reference, \texttt{searchrad}=3. and \texttt{minobj}=6. We drizzled the matched files to obtain the final F125W and F105W images, with the same \textit{astrodrizzle} parameters used for the F814W filter, and \texttt{final\_scale}=0\farcs05, in order to match the pixel scale of the WFC3 images to that of ACS.

In order to check the goodness of our match, we compared the coordinates of sources recovered in all three \textit{HST} final images, considering only the 30 brightest objects. The final mean deviation within the astrometric solutions of the filters is $\sim$0\farcs03. If we compare instead their astrometry with the GAIA DR2 catalog \citep{gaia_collaboration_gaia_2018}, the mean difference in the coordinates of the recovered sources (10 in F814W, 13 in both F105W and F125W) is $\sim$0\farcs05. We note that the final F105W image is affected by an artifact, due to the presence of a satellite trail in one of the \textit{flt} exposure. We decided to not discard this exposure to obtain the deepest image, but caution is needed when examining sources close to the trail. 
The final reduced images F814W, F105W and F125W (hereafter $i_{814}$, $Y_{105}$, and $J_{125}$) are presented in Figure \ref{rgb} as an RGB color image created with JS9-4L (v2.2; \citealt{mandel_ericmandeljs9_2018}).

\subsubsection{Point-Spread Function (PSF) Matching}
Finding high-redshift galaxies requires very accurate colors from photometric measurements in different bands. We calculate the photometry in fixed aperture diameters of 0\farcs4, as later discussed in Section \ref{selection}, and therefore imaging in all bands need to be matched to the same PSF. \edit1{The size in pixels of the PSF in the $i_{814}$, $Y_{105}$, and $J_{125}$ images are 2.6, 4.45, and 4.55, respectively.} The reference matching image is the one with the largest PSF full width at half maximum (FWHM), which in this case is the $J_{125}$ band. 
\replaced{For every band, we first attempted to select appropriate stars contributing to the building of the PSF by making a half-light radius versus magnitude diagram to identify the stellar locus. However, since there are not many stars in the field, and several were saturated, we did not manage to obtain a reliable PSF model from the images.}{We decided against using stars to build the PSF in each band because of their scarcity. Hence,} to perform the PSF matching we therefore relied on the standard \hst\ PSFs produced with a high level of precision by STScI\footnote{\url{https://www.stsci.edu/hst/instrumentation/wfc3/data-analysis/psf}} from the \texttt{\char`_flt / \char`_flc} frame. The matching kernel for image convolution is produced with \texttt{pyPHER} \citep{boucaud_convolution_2016} to make the final PSF-matched images.


\section{Making the Catalogs}\label{photometry}
This analysis follows closely the procedure for Lyman-break detection in \citet{rojas-ruiz_probing_2020}. We utilize the Software \texttt{Source Extractor} v2.25.0 to measure the photometry of the sources in all three \hst\ filters in dual-mode with a coadded $Y_{105}+J_{125}$ as the detection image, which serves to maximize the signal-to-noise (S/N) and minimize the number of spurious sources resulting in the catalogs. The errors provided by \texttt{Source Extractor} depend on the RMS map. We build this RMS map for each band from the sky flux measurements in the science image (SCI) found with a 2.5-$\sigma$ clipping, and the reduced weight image (WHT) as follows: 
\begin{equation}
RMS = \frac{\sigma_{SCI}}{Med\{1/\sqrt{WHT}\}} \times \frac{1}{\sqrt{WHT}}
\end{equation}

The flux of the objects is measured in a small Kron elliptical aperture (PHOT\_AUTOPARAMS 1.2, 1.7) \edit2{which is subsequently corrected} up to total magnitudes using the flux measured in a larger Kron aperture (PHOT\_AUTOPARAMS 2.5, 3.5), as previously done in high-redshift galaxy studies \citep[e.g.,][]{bouwens_discovery_2010, bouwens_new_2021, finkelstein_stellar_2010,finkelstein_census_2022}. To identify point-like sources in our catalog, \replaced{rather than rely on}{we avoid relying solely on} the CLASS\_STAR parameter from \texttt{Source Extractor} that can be misleading when investigating high-redshift sources \citep[see][]{finkelstein_evolution_2015,morishita_bright-end_2018}, we also perform photometry in a \replaced{0\farcs4 circular aperture in diameter}{0\farcs4-diameter circular aperture.} Comparing the ratio between the Kron elliptical aperture and the 0\farcs4 circular aperture sizes helps identify point-like sources such as stars or bad pixels. This circular aperture also serves as a high S/N measurement of the source at the targeted wavelengths and is thus relevant for the S/N cuts in our criteria for selecting candidates at $z\sim7.5$ as described in \S \ref{selection}. 

Upon visual inspection of the segmentation map produced from the \texttt{Source Extractor} run, the combination of parameters DETECT\_THRESH = 1.5 and DETECT\_MINAREA = 7 maximize the number of sources detected while lowering the spurious fraction. 

\begin{figure*}[t!]
\centering
\includegraphics[width=1.0 \linewidth]{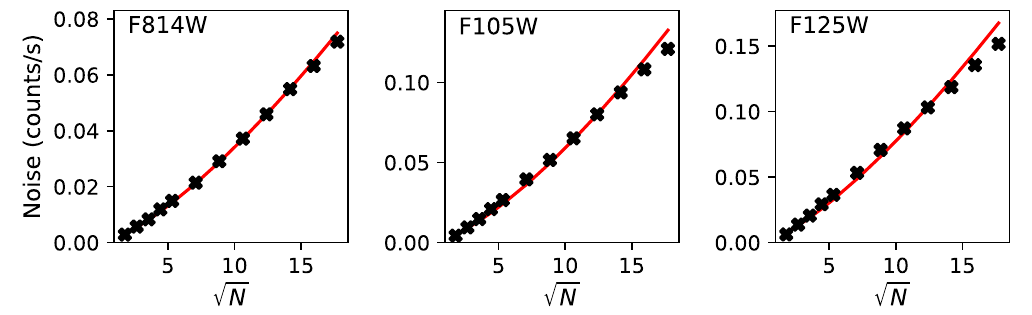}
\caption{ Noise calculation for the field in the three-\hst\ bands following Equation \ref{noise_eq}. $N$ is the number of pixels in the area of the aperture with diameters 2.0, 3.0, 4.0, 5.0, 6.0, 8.0, 10, 12, 14, 16, 18, 20 (pixel scale is 0\farcs05). Note how the noise grows with a bigger aperture, as expected from the equation. The red line shows the best fit correlating the noise and aperture size $N$, which is used to find the $\alpha$ and $\beta$ free parameters that contribute to the noise estimate. }
\label{noise_fig}
\end{figure*}

\subsection{Noise Calculation}\label{noise}
We perform an empirical noise calculation of the images \edit1{to account for the partially correlated noise characteristic of drizzled \hst\ images \citep{casertano_wfpc2_2000}. While \texttt{Source Extractor} calculates the flux uncertainties from individual uncorrelated pixels in the RMS map, the procedure described in \citet{papovich_spitzer-hetdex_2016} accounts for correlated and uncorrelated noise. We closely follow this empirical noise estimate as described below.}

For images with exclusively uncorrelated pixels, the noise is measured in a circular aperture of $N$ pixels which scale following $\sigma_n = \sigma_1 \times \sqrt{N}$, where $\sigma_1$ is the pixel-to-pixel standard deviation of the background. Conversely, the noise from completely correlated pixels is measured as $\sigma_n = \sigma_1 \times N$. In our \hst\ images, the noise truly varies among both correlations as $N^\beta$ where $0.5<\beta<1$, and this can be estimated for the whole image with the parameterized equation:

\begin{equation}
\sigma_n = \sigma_1(\alpha N^\beta),
\label{noise_eq}
\end{equation}

where $\alpha$ has to be a positive value. Note that we do not include the Poisson correction of the equation from \citet{papovich_spitzer-hetdex_2016} as it did not contribute to the calculation of the noise in the \hst\ images. We measure the noise in each of the three \hst\ images by first placing randomly-distributed apertures in the sky background with growing sizes from 0\farcs1 to 1\farcs0 in diameter. Then, we use the \texttt{curve\_fit} Python function with the Levenberg-Marquardt least-squares method to fit for the noise found in the random apertures with increasing size (see Figure \ref{noise_fig}). The calculated noise values in each image are applied to \replaced{the Kron and 0\farcs4 apertures to calculate the flux error for these aperture sizes.}{estimate the flux errors for the Kron and 0\farcs4 apertures.} \edit1{For the Kron aperture, $N$ is calculated as the number of pixels in the ellipse defined by the semi-major (\texttt{A\_IMAGE}) and semi-minor (\texttt{B\_IMAGE}) axes measured by \texttt{Source Extractor}. }  

\subsection{Corrections to the Photometry Catalogs}
The resulting \texttt{Source Extractor} catalogs with the calculated flux errors are then corrected for Galactic dust attenuation following the \citet{cardelli_relationship_1989} extinction curve with an $R_v = 3.1$, as motivated in similar studies of high-redshift galaxies \citep[e.g.,][]{rojas-ruiz_probing_2020,finkelstein_census_2022, tacchella_stellar_2022}. \edit1{We use \citet{schlafly_measuring_2011} to correct for galactic extinction and find a color excess $E(B-V)=0.025$.} The zero points for the final catalog are calculated according to the newest 2020 \hst\ photometric calibrations for ACS and WFC3, which apply to the observed dates of the images. The zero points in AB magnitude are 25.9360, 26.2646, and 26.2321 for $i_{814}$, $Y_{105}$, and $J_{125}$, respectively. We also apply in all filters an aperture correction from the large (2.5, 3.5) to small (1.2, 1.7) Kron aperture photometry measured in the $J_{125}$, to account for the missing PSF flux in the smaller aperture. 

\subsection{\emph{Spitzer}/IRAC Photometry}\label{irac}
Additional \textit{Spitzer}/IRAC 3.6$\mu$m and 4.5$\mu$m images covering the same area of the quasar environment explored with \hst\ are available from cycle 16 archival database (PI: Decarli). \edit1{Each of the IRAC mosaics has an exposure time of 3.4 hrs, and the 3$\sigma$ limiting depth for point sources is $\approx0.8~\mu\mathrm{Jy}$ in both Channels 1 and 2.} These additional photometric bands provide crucial information to distinguish true high-redshift galaxies from lower-redshift contaminants. These IRAC bands allow for better building of the spectral energy distribution (SED), and hence to better differentiate between a dusty Balmer-break galaxy at $z\sim2$ and a high-redshift candidate of interest at $z\sim7.5$. \edit1{The FWHM of the IRAC point response function (PRF) is $\approx 1\farcs8$ in Channels 1 and 2\footnote{\url{https://irsa.ipac.caltech.edu/data/SPITZER/docs/irac/iracinstrumenthandbook/5/}}; this is about two orders of magnitude larger than in \hst.}
Therefore, in order to match the sources from \replaced{both data}{the two data sets} the IRAC PSFs need to be modeled in order to correct for deblending of sources and calculate accurate flux and flux errors. The mosaics and modeling are performed following \cite{kokorev_alma_2022}, and are briefly described here. 

To obtain photometry from the IRAC imaging, a PSF model method is produced using the tools from the Great Observatories Legacy Fields IR Analysis Tools \citep[GOLFIR;][]{brammer_gbrammergolfir_2022}. This modeling method uses a high-resolution prior, which is built from combining the \hst/ACS and WFC3 images. This resulting image is combined with the IRAC PSF using a matching kernel to finally obtain the low-resolution templates. The original IRAC images are divided into homogeneous $4\times4$ patches of 120\farcs0, which are allowed to overlap to improve the modeling. The brightest stars and sources with high signal-to-noise ratios in the IRAC and \hst\ images are manually masked to avoid large residuals from the fit. IRAC model imaging are first generated for the brightest objects in the $J_{125}$ catalog doing a least-squares fit of the low-resolution IRAC patches to the original IRAC data to obtain the modeled fluxes. The flux errors are simply the diagonal of the covariance matrix of the model. Similarly, for the fainter sources in the catalog but the least-squares fit normalizations are then adopted as the IRAC flux densities. The resulting photometry from this PSF modeling method is used for the rest of the analysis.

\section{Selection of Galaxy Candidates}\label{selection}
Galaxy candidates neighboring the quasar \Pisco\ are found with a similar method as previous work in the literature \citep{rojas-ruiz_probing_2020, finkelstein_census_2022,bagley_bright_2024}. We rely on the photometric redshift technique by fitting the best SED model to the \hst\ and \emph{Spitzer} photometry. We \replaced{constrain}{refine} the catalog of candidates by applying S/N cuts, quality checks between the low and high-redshift fitting, and color--color comparisons to low-redshift interlopers and MLT brown dwarfs (see \S\ref{z7p5}). The different steps to obtain the catalog of galaxy candidates are described in this section.

\subsection{Photometric Redshifts with EAZY}
We use the ``Easy and Accurate Z$_{phot}$ from Yale" (\texttt{EAZY}; \citealt{brammer_eazy_2008}) version 2015-05-08 to calculate the photometric redshifts of all sources in our catalogs. \texttt{EAZY} calculates the probability distribution function of photometric redshifts $P(z)$ based on a minimized $\chi^2$ fit of the observed photometry in all given filters to different SED models of known galaxy types. \texttt{EAZY} includes the 12 \texttt{tweak\_fsps\_QSF\_12\_v3} Flexible Stellar Population Synthesis (FSPS) models \citep{conroy_propagation_2009,conroy_propagation_2010}, the template from \citet{erb_physical_2010} of the young, low-mass and blue galaxy BX418 at $z = 2.3$ exhibiting high equivalent-width (EW) of nebular lines and \lya, and a version of this galaxy without the \lya\ emission to mimic  attenuation from the intergalactic medium (IGM) while preserving strong optical emission lines. All these 14 templates are fed equally into \texttt{EAZY} so that it constructs the best-fitting models from a linear combination of the templates to the flux and flux errors of the source measured in the Kron 1.2, 1.7 elliptical aperture (see Section \ref{photometry}). For each source, \texttt{EAZY} applies IGM absorption following \cite{inoue_updated_2014} for redshift steps of $\Delta z = 0.01$. Initially, we consider the redshift probability distribution when giving the templates freedom from $z=0.01 - 12$ and assume a flat luminosity prior, as galaxy colors at $z \gtrsim 6$ are not yet well understood \citep{salmon_relics_2018}. This wide redshift range is chosen \replaced{so that we can evaluate the probability for galaxies truly neighboring \Pisco\ at $z \sim 7.5$ and their corresponding probability of being a contaminant source at lower redshift ($z<5$).}{to allow the comparison between the probability of galaxies to be at high- ($z>5$) and low- ($z<5$) redshifts.}

\subsection{Selection Criteria for Catalog}\label{criteria}
We build the final catalog of galaxy candidates \replaced{from the results of the photometric redshift fits from \texttt{EAZY} and applying the following selection criteria.}{applying the following selection criteria to the results of the photometric redshift fits from \texttt{EAZY}:}
\begin{itemize}
\item $S/N_{i_{814}} < 2.0$ measured in the 0\farcs4 circular aperture, implying a non-detection in $i_{814}.$

\item $S/N_{Y_{105}}$ or $S/N_{J_{125}} > 5.0$ also measured in the 0\farcs4 circular aperture, to ensure the source is detected at high redshift while also potentially selecting strong Lyman-$\alpha$ emitters where the flux would only be detected in the $Y_{105}$, or galaxies with strongly absorbed Lyman-$\alpha$ producing continuum emission only in the $J_{125}$.

\item The integrated redshift probability $P(z)$ calculated from \texttt{EAZY} at $P(6<z<12) > 60\%$, securing that a high-redshift solution dominates over the total probability distribution. 

\item The integral of the primary peak of the total integrated distribution $P(z_{peak}) > 50\%$.

\item The redshift probability distribution at $z=7.5$ is higher than the neighboring distributions, in $\Delta z=1$ bins:\

\edit1{$P(6<z<7) < P(7<z<8)\  \wedge$ \\
$P(8<z<9) < P(7<z<8)$}

\end{itemize}

\begin{figure*}[ht!]
\centering
\gridline{\fig{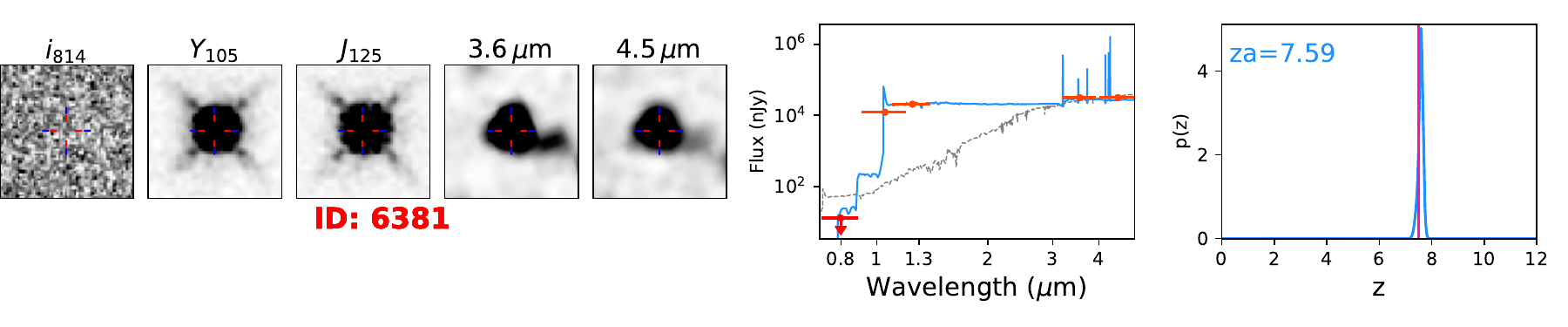}{\textwidth}{}}
\vspace{-16mm}
\gridline{\fig{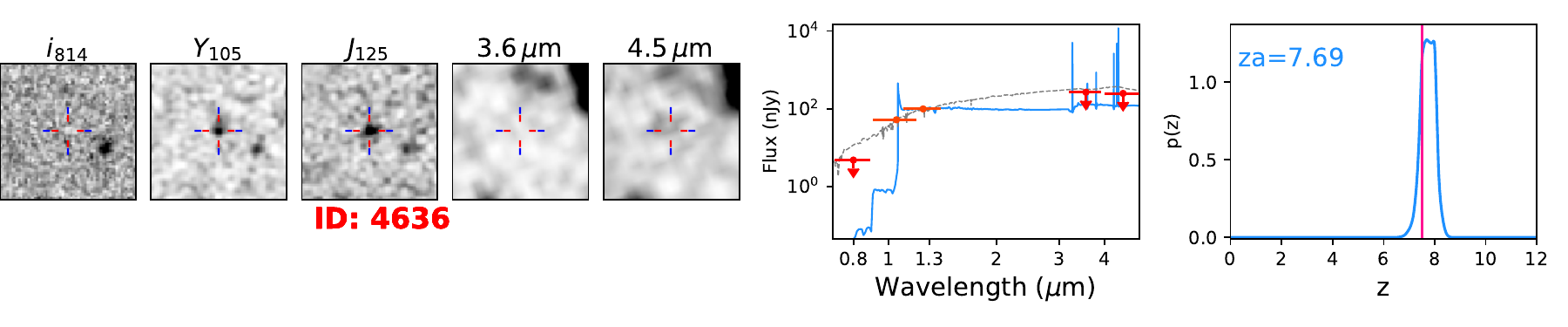}{\textwidth}{}}
\vspace{-16mm}
\gridline{\fig{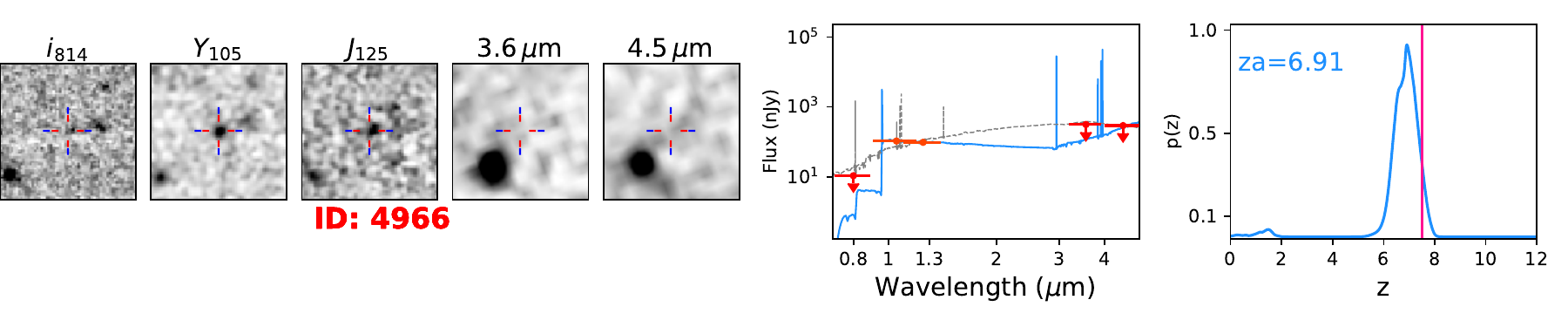}{\textwidth}{}}
\vspace{-16mm}
\gridline{\fig{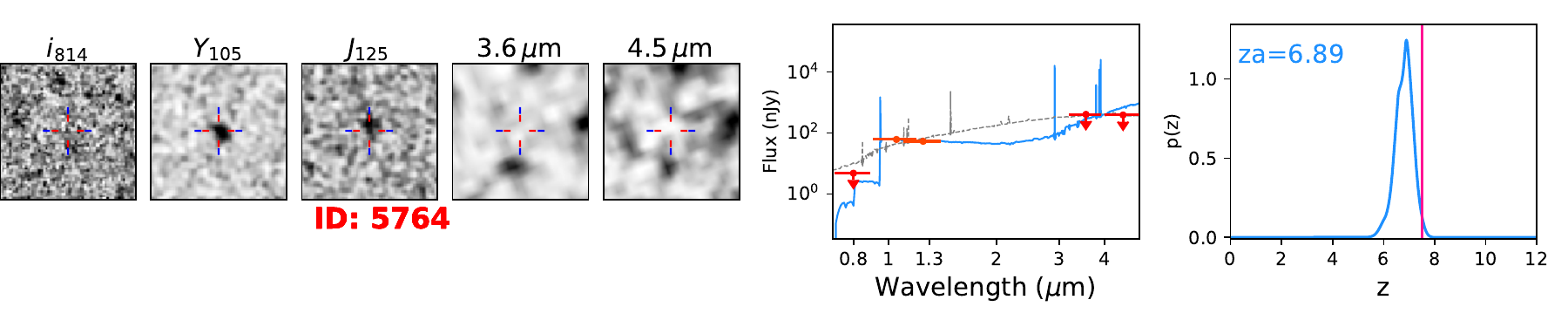}{\textwidth}{}}
\vspace{-10mm}
\caption{Galaxy candidates resulting from our search including the quasar \Pisco\ at $z=7.54$, and a new LBG candidate in its environment. \textit{Left:} Postage stamps of each candidate in the $iYJ$ {\it HST} filters (3$\farcs0 \times3\farcs$0), and the two \spitzer/IRAC bands (12$\farcs0 \times 12\farcs$0). {\it Middle:} The best-fit SED of the high-redshift solution of the candidate is presented in blue, with non-detections in red as $1\sigma$ upper limits. We present the SED of the low-redshift solution in a dotted grey line. {\it Right:} The $P(z)$ versus $z$ from \texttt{EAZY} with the best-fitting redshift (za) in blue, and a vertical pink line indicating the redshift of the quasar for reference. Note that the redshift probability distributions are highly favored at $z=7-8$ for the top two panels, where the quasar is ID: 6381. The LBG candidate C-4636 has a slightly higher redshift solution at $z_{phot}=7.69$ because of the flat $P(z)$ across $z=7.5-8$. The bottom two candidates favor a slightly lower redshift solution at $z\sim7$.}
\label{cand1}
\end{figure*}

We do not place a cut in the half-light radius of the source in order to include in the catalog possible active galactic nuclei (AGN) sources, which would exhibit a more point-like morphology. However, this parameter is reported in Table \ref{table-phot} and is considered during the visual inspection step. Note that the half-light radius $r_{0.5}$ of a star in our survey in the $J_{125}$ band is 2.65 pixels, or 0\farcs13.

Using the above criteria we find 5 LBG candidates where one is the quasar \Pisco, and two are identified as diffraction spikes from visual inspection. The succeeding catalog is thus composed of the recovered quasar and two galaxy candidates. We note that decreasing the S/N threshold so that $S/N_{Y_{105}}$ or $S/N_{J_{125}} > 3.0$  results in \replaced{contamination from marginal detection sources }{a large contamination due to sources with a marginal detection} in just one band (25), diffraction spikes (13), bad pixels or other detector artifacts (25).

An additional test fitting only a lower-redshift solution was performed to better discriminate among possible low-redshift contaminants. For this, we set \texttt{EAZY} to freely fit the 14 SED templates over a redshift span $z = 0.01-5$. We then compared the $\chi^2$ of the best-fit template from this lower-redshift solution $\chi^2_{lowz}$ \edit1{to that at higher-redshift $\chi^2_{highz}$ with the redshift span $z = 0.01-12$.} If $\Delta{\chi^2_{l-h}} =\chi^2_{lowz} - \chi^2_{highz} < 4$ the goodness of the fit is lower than the threshold of 95\% confidence interval, which means the source can be similarly fit with a high and a lower redshift solution \citep[see e.g.][]{finkelstein_census_2022,bagley_bright_2024}. We discarded one candidate \replaced{not passing this test having}{that did not pass this test, as it had} a $\Delta{\chi^2_{l-h}} = 2.4$ with a redshift solution $z_{low}=2.5$ and $z_{high}=7.9$. The final catalog thus contains the quasar and one LBG candidate at $z\sim7.5$ passing the test with $\Delta{\chi^2_{l-h}} = 16.51.$

\begin{figure*}[ht!]
\centering
\gridline{\fig{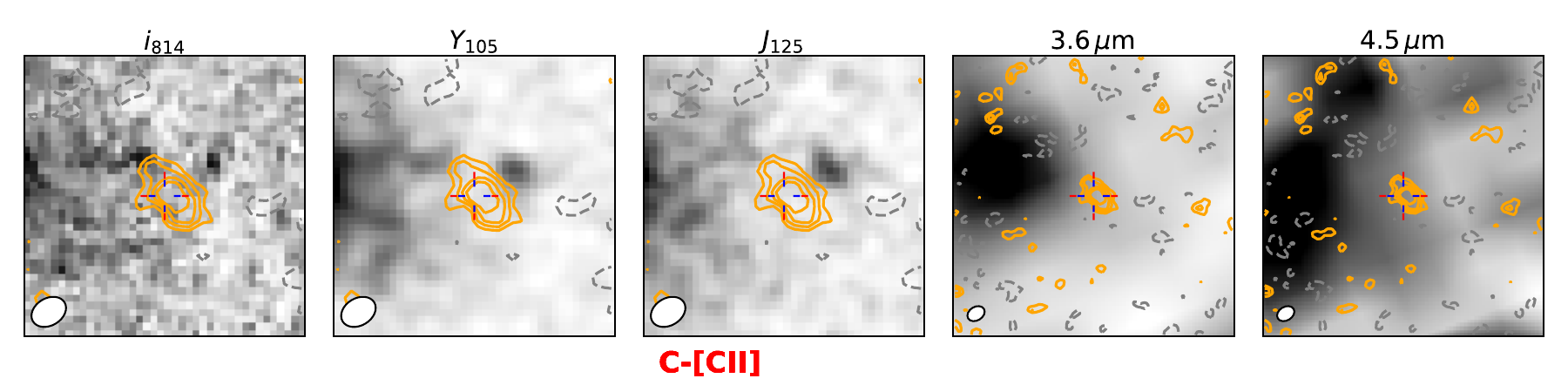}{\textwidth}{}}
\vspace{-9 mm}
\caption{Postage stamps in the \hst\ (2\farcs0 $\times$ 2\farcs0) and \spitzer\ (4\farcs0 $\times$ 4\farcs0) filters used in this work at the position of the dusty star-forming galaxy (DSFG) previously identified with ALMA at $z = 7.5341\pm0.0009$ \citep{venemans_kiloparsec-scale_2020}. The rest-frame \cii-158$\micron$ emission is presented with the contours at levels ($-2$, $2$, $3$, $4$, $5$)$\sigma$ where the rms value $\sigma=0.04$ mJy~beam$^{-1}$. The ALMA 222.7\,GHz beam with size 0\farcs26 $\times$ 0\farcs19 is shown in white at the bottom-left corner. This \cii-emitter is not detected in any of the near-IR filters we use.}
\label{cand-cii}
\end{figure*}

\begin{deluxetable*}{lccccccccc}[ht]
\tablecaption{EAZY fit of Galaxy Candidates in the Quasar Field\label{table-pz}}
\tablewidth{700pt}
\tabletypesize{\scriptsize}
\tablehead{
\colhead{ID} & 
\colhead{$\alpha$} & 
\colhead{$\delta$} & 
\colhead{$P(z>6)$} &
\colhead{$P(6.5<z<7.5)$} &
\colhead{$P(7<z<8)$} &
\colhead{z$_{phot}$} &
\colhead{$\Delta{\chi^2_{l-h}}$} &
\colhead{z$_{spec}$} &
\colhead{$d$}
\\
\colhead{} &
\colhead{(deg)} & 
\colhead{(deg)} & 
\colhead{\%} & 
\colhead{\%} & 
\colhead{\%} & 
\colhead{} & 
\colhead{} & 
\colhead{} & 
\colhead{(")}
}
\startdata
QSO--6381 & 205.5337428 & 9.4773167 & 100 & 17 & 100 & 7.59$^{+0.08}_{-0.11}$ & 483.16 & 7.5400$\pm$ 0.0003\tablenotemark{$\dag$} & {}\\
C--4636 & 205.5435057 & 9.4851103 & 100 & 20 & 80 & 7.69$^{+0.33}_{-0.23}$ & 16.51 & ... & 44.6\\
C--\cii\ & 205.5343208 & 9.4787250 & ... & ... & ... & ... & ... & 7.5341$\pm$0.0009\tablenotemark{$\ast$} & 5.44\\
\hline
\hline
C--4966 & 205.5207121 & 9.4853401 & 95 & 71 & 39 & 6.91$^{+0.41}_{-0.51}$ & 6.48 & ... & 90.4\\
C--5764 & 205.5591434 & 9.4785146 & 98 & 79 & 29 & 6.89$^{+0.26}_{-0.43}$ & 14.05 & ... & 55\\
\enddata
\tablecomments{This table presents the catalog of galaxy candidates in the quasar field, selected here with \texttt{EAZY} using \hst\ and \textit{Spitzer} photometry. The reported values at the top correspond to the recovered quasar \Pisco\ with confirmed systemic redshift $z=7.54^\dag$, candidate C--4636 at $z\sim7.5$, and dust-obscured candidate C-\cii\ identified with a systemic redshift ${z_{\mathrm{[C\,{\sc II}]}}}^\ast$. For the bottom two candidates, our fit preferred a solution at $z\sim6.8-6.9$, in the redshift range of an absorber at $z=6.84$ in the line-of-sight of the quasar \citet{simcoe_interstellar_2020}. Column 1 is the candidate ID. Columns 2-3 are the RA and DEC calculated in degrees. Columns 4-6 present the integral of the redshift probability distribution in three redshift bins (see \S \ref{criteria}). Column 7 presents the photometric redshift with the highest probability and its 68\% confidence interval as calculated with \texttt{EAZY}. Column 8 shows the difference in the best-fit ${\chi^2}$ for the low- ($z<5$) and high- ($z<12$) redshift solutions. Column 9 is the spectroscopic redshift of the sources when available. Column 10 is the projected distance of the candidate to the quasar.}
\tablenotetext{$\dag$}{Systemic redshift measured using ALMA observations of the \cii$-158\micron$ emission line from the quasar's host galaxy in \citet{banados_z_2019}}
\tablenotetext{$$\ast$$}{Systemic redshift calculated from \cii$-158\micron$ observations with ALMA in \citet{venemans_kiloparsec-scale_2020}}
\vspace{-9mm} 
\end{deluxetable*}

\section{Galaxy Candidates in the Quasar Field}\label{results}
In this section, we present the results from our search of galaxy candidates associated with the quasar \Pisco\ environment at $z\sim7.5$. We further comment on
the inspection of our \hst+\textit{Spitzer}/IRAC data at the position of gas-rich \cii-emitter at $z\sim7.5$ previously identified in \citet{venemans_kiloparsec-scale_2020}. Finally, we explore additional galaxy candidates at a slightly lower redshift than that of the quasar, at $z\sim7$. Figure \ref{cand1} shows the postage stamps of the LBG candidates, their SED, and photometric redshift solution from \texttt{EAZY} with both the high-redshift and lower-redshift fits. Table \ref{table-pz} summarizes the \replaced{catalog}{properties} of these LBGs and \cii-emitter.

\begin{deluxetable*}{lcccccccccccc}[ht]
\tablecaption{Photometry of \hst\ and \spitzer\ Selected Galaxy Candidates \label{table-phot}}
\tablewidth{700pt}
\tabletypesize{\scriptsize}
\tablehead{
\colhead{ID} & 
\colhead{S/N$_{i}$} & 
\colhead{S/N$_{Y}$} &
\colhead{S/N$_{J}$} &
\colhead{S/N$_{3.6\micron}$} &
\colhead{S/N$_{4.5\micron}$} &
\colhead{$i_{814}$} & 
\colhead{$Y_{105}$} & 
\colhead{$J_{125}$} & 
\colhead{${3.6\micron}$} &
\colhead{${4.5\micron}$} &
\colhead{$r_{0.5}$} 
\\
\colhead{} & 
\colhead{} & 
\colhead{} & 
\colhead{} & 
\colhead{} & 
\colhead{} & 
\colhead{(AB mag)} &
\colhead{(AB mag)} & 
\colhead{(AB mag)} & 
\colhead{(AB mag)} & 
\colhead{(AB mag)} & 
\colhead{(arcsec)}
}
\startdata
QSO--6381 & -0.01 & 3461.1 & 3772 & 100.87 & 118.14 & $>28.62$ & 21.18 & 20.61 & 20.16 & 20.15 & 0\farcs14\\
C--4636 & -1.20 & 8.63 & 12.03 & -1.0 & 0.09 & $>28.71$ & 27.11 & 26.41 & $>24.14$ & $>24.22$  & 0\farcs10\\
\hline
\hline
C--4966 & 1.88 & 9.45 & 6.62 & 0.26 & 0.98 & $>28.77$ & 26.34 & 26.43 & $>23.96$ & $>24.04$ & 0\farcs14\\
C--5764 & 0.24 & 10.64 & 6.16 & 0.90 & 0.19 & $>28.80$  & 26.93 & 27.08 & $>23.68$ & $>23.73$ & 0\farcs15\\
\enddata
\tablecomments{This table presents the photometry of the \replaced{catalog of galaxy candidates}{high-redshift galaxy candidates.} Column 1 is the candidate ID. Columns 2-6 are the calculated signal-to-noise values from the 0$\farcs$4-diameter circular aperture in the {\it HST} bands, and from the \spitzer\ photometry. Columns 7-11 are the calculated AB magnitudes, where the limiting magnitudes correspond to 3$\sigma$ estimates. Column 12 is the half-light radius of the object in arcseconds.}
\end{deluxetable*}

\subsection{A galaxy candidate at $z\sim7.5$}\label{z7p5}
We recover the quasar with a photometric redshift of $z_{phot}=7.59$, where its systemic redshift measured from
\cii\ emission is $z =7.5400 \pm 0.0003$ \citep{banados_z_2019}. We find a new LBG candidate, C-4636, at $z_{phot}=7.69$ (see Figure \ref{cand1}). This photometric redshift is slightly higher than \replaced{$z=7.5$}{that of the quasar} given the very flat redshift probability distribution between $z=7.5-8$, but the solution dominates among the other redshift distributions with $P(7<z<8)= 80\%$, constraining its association with the environment of the quasar. Moreover, C-4636 is at a projected distance of 223 pkpc from the quasar. \edit1{This distance is similar to that of galaxies found around other high-z quasars in the literature, which showed strong quasar-galaxy clustering} \citep[e.g.,][]{morselli_primordial_2014,farina_mapping_2017,mignoli_web_2020,meyer_constraining_2022}. \edit1{Further exploration of the QSO-LBG clustering for quasar \Pisco\ and this candidate is described in \S\ref{overdensity}.}

\edit1{We examine the candidate's  $i_{814}-Y_{105}$ and $Y_{105}-J_{125}$ colors to evaluate possible stellar contamination. We take MLT-dwarf stars from the IRTF SpeX Library developed by \citet{burgasser_spex_2014} and compare their colors to those of LBGs and quasars at $z=6.5-8.5$. The color of candidate C-4636 in} $Y_{105}-J_{125} = 0.7$ and its compact morphology with a \edit1{half-light radius $r_{0.5} = 0\farcs10$ is comparable with} the average radius of stars in this field \edit1{of $r_{0.5} = 0\farcs13 \pm 0\farcs01$}, make it a possible MLT-dwarf contaminant (see Figure \ref{color-color}). However, the ratio of the flux in the total Kron to the $0\farcs4$ aperture \edit1{of $1.45\pm0.07$}, and the \texttt{Source Extractor} stellarity parameter of CLASS\_STAR $=0.08$ do not classify this source as a star. Additionally, galaxies at this redshift would show a distinct SED from MLT contaminants at $\lambda >2 \mu$m, which can be determined even with shallow IRAC imaging \citep[e.g][]{finkelstein_census_2022, bagley_bright_2024}. The non-detection in our IRAC/3.6$\mu$m and 4.5$\mu$m imaging support the high-redshift nature of this candidate. \replaced{Thus, we include this source as an LBG candidate in the physical environment of the $z=7.54$ quasar \Pisco. However, photometry in contiguous filters redder of the $J_{125}$ band would provide better color diagnostics.}{Thus, we still consider this source as a good LBG candidate in the physical environment of \Pisco. We recognize that additional filter information redder than the $J_{125}$ band would provide further insights into the characterization of this source.}

The compact morphology could suggest that C-4636 is an AGN. This is consistent with theoretical simulations which show that AGNs tend to cluster near quasars with SMBH $10^8 - 10^9$\Msun \citep{costa_environment_2014}, and some observational cases already seen \edit1{at $z>5$ \citep[e.g.][]{mcgreer_constraint_2016,connor_x-ray_2019, yue_candidate_2021,maiolino_jades_2023,scholtz_gn-z11_2023}}. Existing 45ks \textit{Chandra} observations of this field do not show any X-ray signal at this location \citep{banados_chandra_2018}. Future JWST NIRSpec spectrum targeting strong nebular emission lines could be used for a Baldwin, Philips, \& Terlevich (BPT) diagnostic diagram \citep{baldwin_classification_1981} to distinguish whether this source is an AGN.

\subsection{Dusty Star-Forming Galaxy}
A galaxy candidate at 27 projected pkpc from the quasar \Pisco\ had been previously identified in 
\cite{venemans_kiloparsec-scale_2020}. This candidate was recovered with ALMA observations targeting the rest-frame \cii-158$\mu$m emission of the quasar and found to be at $z_{\mathrm{[C\,{\sc II}]}}=7.5341\pm0.0009$. In order to study the properties of this source in other wavelengths, we looked for any counterpart emission in the \hst\ and \spitzer/IRAC data. However, we did not detect this galaxy in any of the five images (see Figure \ref{cand-cii}). Since no flux is recovered up to SNR $\sim2$, the possibility of this galaxy being a low-redshift interloper is strongly disfavored. Furthermore, many studies have concluded that there is a significant population of dust-obscured galaxies that have no rest-frame optical counterpart detected at the current observational limits \citep[e.g.][]{mazzucchelli_spectral_2019, wang_dominant_2019, meyer_constraining_2022}. Therefore, this \cii-emitter could be one of these dusty star-forming galaxies (DSFG) in the environment of \Pisco. Currently, the only detection available of this galaxy is from ALMA 223\, GHz observations of its \cii\ emission at $0.10\pm0.03$ Jy km~s$^{-1}$, and no dust continuum \added{was recovered} \replaced{(upper limit, $<0.06$)}{\citep[$f_{\mathrm{cont}}$ $<0.06$ mJy;][]{venemans_kiloparsec-scale_2020}}. \replaced{Thus we cannot model a good SED of this galaxy to predict its emission at other wavelengths.}{Given the information at hand, we are not able to place meaningful constraints or predictions on the SED of this source.} Further ALMA or \jwst\ observations are necessary to confirm the nature of this source and study its properties. The current ALMA observations of this field cover only a FOV $\sim39"$, i.e. $\sim194$ pkpc, around the quasar, hence we are not able to investigate any counterpart for the LBG candidates found in this work.

\subsection{Additional Candidates at $z\sim7$}
\begin{figure}
\centering
\includegraphics[width=\linewidth]{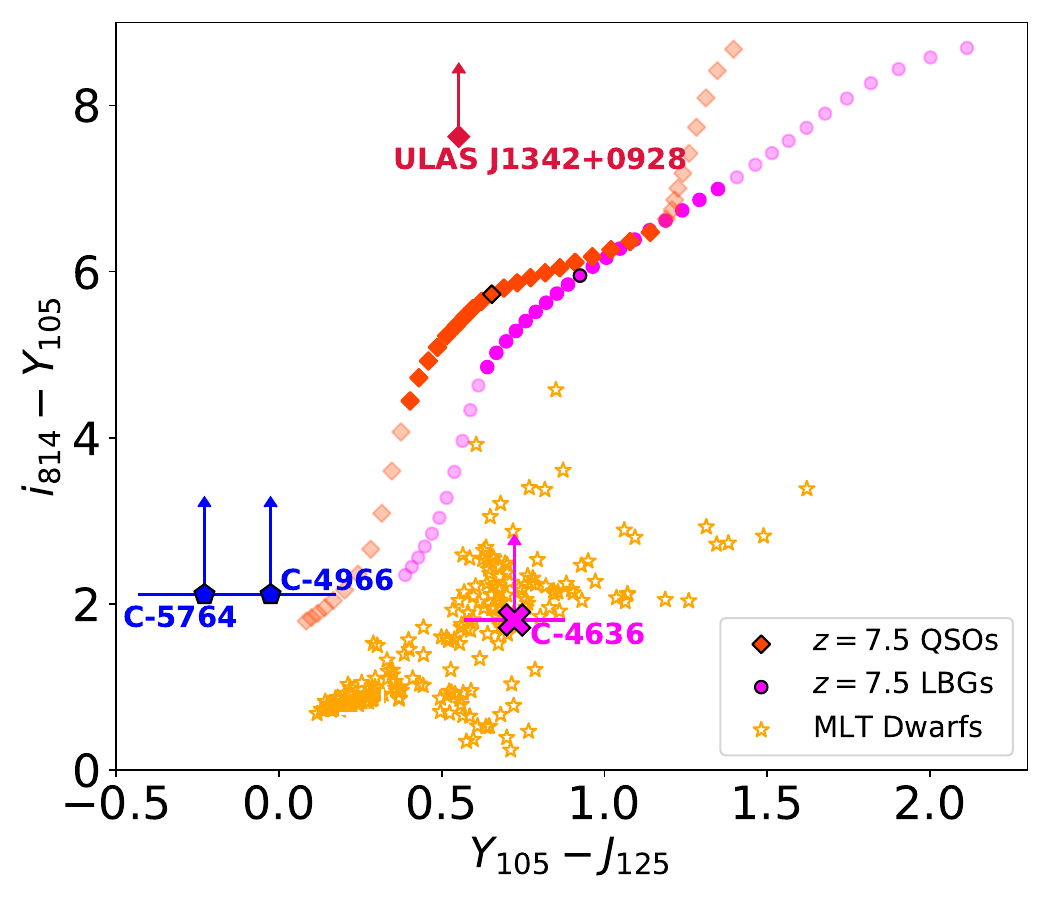}
\caption{\edit1{Color--Color diagram using the \hst\ bands. The quasar \Pisco\ is marked with a crimson diamond, indicating a strong Lyman Break in the $i_{814}-Y_{105}$ color. 
We show typical quasars (red diamonds) and LBGs (magenta circles) colors in the redshift range $ z =  6.5 - 8.5$, with a redshift step $\Delta_z = 0.05$. The MLT-dwarf stars are denoted by yellow stars. LBG candidate C-4636 identified at $z_{phot}=7.69$ with \texttt{EAZY}, is marked with a magenta cross. Blue pentagons represent the $z\sim7$ LBG candidates, showing distinctively bluer \edit2{$Y_{105}-J_{125}$} colors compared to $z\sim7.5$ LBGs in the diagram.}}
\label{color-color}
\end{figure}

\citet{simcoe_interstellar_2020} inspected the optical/NIR spectrum of \Pisco, and identified a strong metal absorber at $z=6.84$ \edit1{spanning $\sim150$~km~s$^{-1}$} on its line-of-sight. This galaxy has not been directly observed in emission yet. Hence, we also explore candidates at $z\sim7$ to search for any counterparts \citep[e.g.][]{neeleman_c_2019}. \replaced{For this we select}{We begin by selecting} all sources with $P(6<z<12) > 60\%$, and a $P(6<z<7)$ higher than neighboring distributions in $\Delta z=1$ bins. After visually inspecting the candidates and evaluating the best fits for the high and lower-redshift solutions $\Delta{\chi^2_{l-h}}$, \replaced{as we did for}{similar to the process used for} the $z\sim7.5$ galaxy candidates search, we identify two galaxy candidates. C-4966 has a photometric redshift of $z_{phot}=6.91$ and a probability distribution $P(6<z<7)=55\%$ and $P(6.5<z<7.5)=71\%$. C-5764 is found at $z_{phot}=6.89$ and has a $P(6<z<7)=68\%$ and $P(6.5<z<7.5)=79\%$ (see Figures \ref{rgb}, \ref{cand1}, and Table \ref{table-pz}). Both candidates favor a redshift solution closer to $z=6.5-7.5$. \edit1{ Evaluating different Ly-$\alpha$ properties for these candidates using our color models for LBGs at $z=6.5-8.5$ in Figure \ref{color-color}, we find that similar colors would be reproduced with either (FWHM$ =1000$ \AA; EW $=15 $\AA) or (FWHM$ = 2000$ \AA; EW$ = 15$ \AA), rather than a more typical narrow Ly-$\alpha$ line for LAEs (FWHM$ = 200$ \AA\ and  EW $=100 $\AA\ or lower). Spectroscopically confirming these galaxies could point to} galaxy clustering in the field at $z\sim6.8$, near the absorber.\\

\section{Discussion}\label{discussion}
\subsection{Completeness}\label{completeness}
\replaced{LBG candidates at $z\sim7.5$ have the Lyman-break at $\sim1\,\mu$m,}{The Lyman break of LBG candidates at $z\sim7.5$ falls at the observed wavelength $\lambda\, \sim 1\mu$m}, which is positioned right in the middle of the $Y_{105}$ band used in this work (see Figure \ref{filters}). This implies that even if the source is detected in $Y_{105}$, its $Y_{105}-J_{125}$ color would be red. No contiguous filter (e.g. \hst\ $JH_{140}$ or $H_{160}$) \added{is available to us }to robustly measure the rest-frame UV color redward of the Ly-$\alpha$ break. Although our photometry in the \textit{Sptizer}/IRAC 3.6 $\mu$m and 4.5 $\mu$m bands complement the galaxy SED and help determine its dust components to rule out low-redshift contaminants, this is only possible for galaxies already robustly selected at high-redshift with \hst\ imaging up to $\lambda \sim 2 \mu$m. Our data currently has a wider gap in the wavelength coverage ($1.4 \mu$m $ - 3.6 \mu$m, or $J_{125} - $ IRAC/$3.6 \mu$m), causing a considerable number of galaxies to never enter the selection catalog. Taking into account all these limitations, we compare the found number density (one new LBG) to what we would recover with a consistent selection of galaxy candidates in the same redshift span of $z\sim 7-8$, using comparable data from blank fields.

The Cosmic Assembly Near-Infrared Deep Extragalactic Legacy Survey (CANDELS) GOODS North and South Deep survey presented in \citet{finkelstein_evolution_2015} provides a similar filter coverage ($I_{814}, Y_{105}, J_{125}, H_{160}$) at comparative depths to our \hst\ survey around \Pisco. This filter coverage helps to assess the completeness of our data since we can attempt to reproduce the number density of LBGs in a blank field, but using just the $I_{814}, Y_{105},$ and $J_{125}$ bands and our selection criteria for high-redshift galaxies. We build the photometry catalog for \texttt{EAZY} using the fluxes from the GOODS catalog, and the flux errors \edit1{ from our limiting magnitudes in the \hst\ filters to match the noise to that of our images.} We run \texttt{EAZY} in the same setup and select LBG candidates with the criteria employed for the analysis of the quasar field (see \S \ref{criteria}). 
Using these three \hst\ filters alone we recover 31 sources. This is much lower than the 125 sources with a photometric redshift between $z=7-8$ in the GOODS catalog \replaced{when including}{that are selected when including} the additional photometry in $H_{160}$. Hence, we recover in total only 31/125, i.e. $\sim25\%$ of the sources.

\begin{figure}[pt!]
\centering
\includegraphics[width=\linewidth]{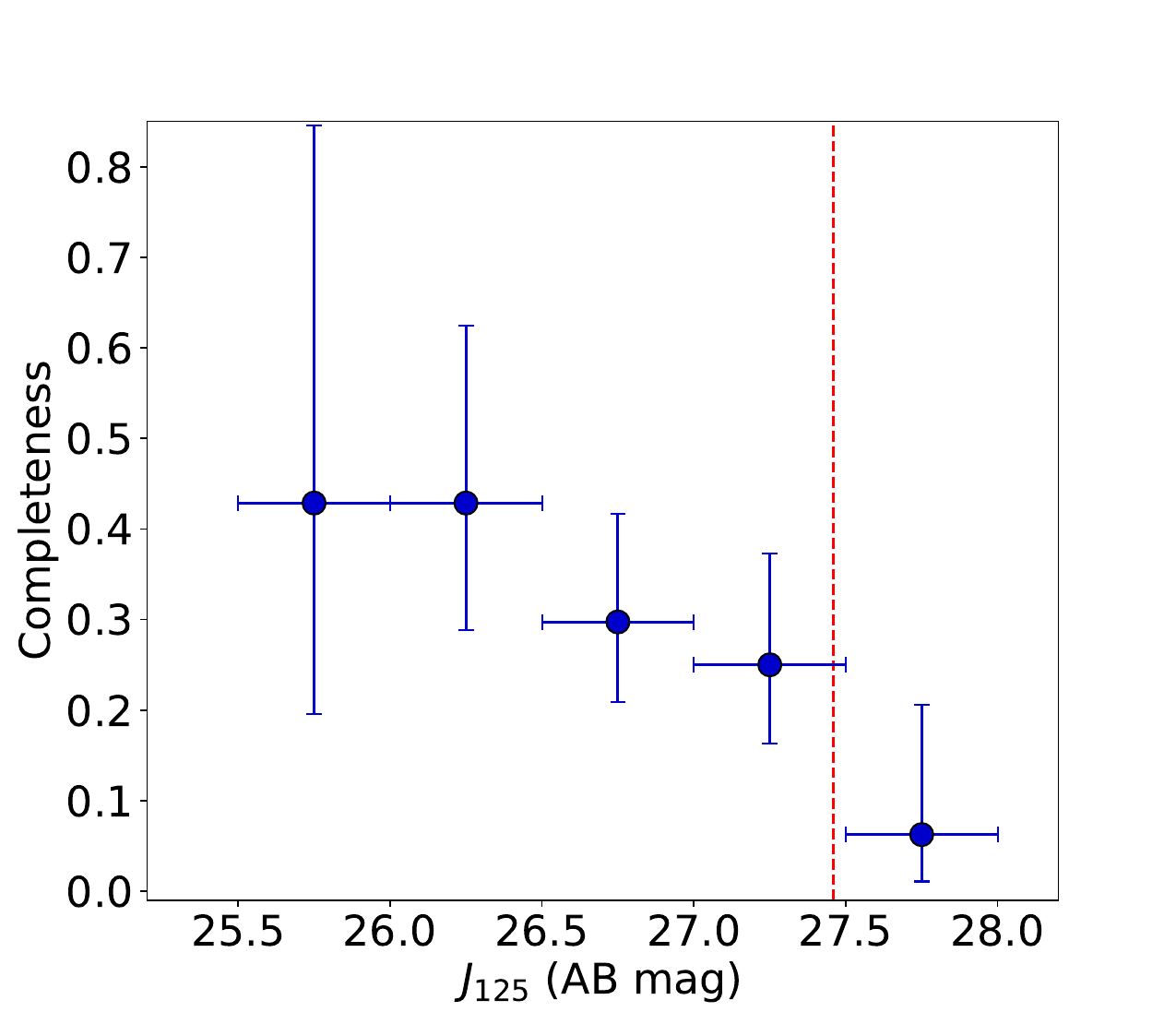}
\caption{Assessment of the completeness of our selection technique of high-redshift galaxy candidates around \Pisco\ presented in magnitude bins of $\Delta_{J_{125}}=0.5$ and 1$\sigma$ Poisson errors calculated with \citet{gehrels_confidence_1986}. \edit1{The $5\sigma$ $J_{125}$ limiting magnitude from this work is denoted with the dashed red line.}}
\label{fig_completeness}
\end{figure}

\replaced{This test demonstrates the incompleteness at recovering $z\sim7.5$ galaxies around \Pisco\ using the currently available filter set.}{This test shows that the selection of $z\sim7.5$ galaxies based on our HST filter set is strongly incomplete.} The recovered fraction of galaxies and their magnitudes in $J_{125}$ is shown in bins of $\Delta 0.5\ mag$ in Figure \ref{fig_completeness}. Note that the faintest galaxy candidate recovered from this catalog has $J_{125}=27.53$, whereas the GOODS catalog from \citet{finkelstein_evolution_2015} has sources as dim as $J_{125}=28.76$. \replaced{The highest recovered rate at 43\% of galaxy candidates from GOODS are those with $J_{125} < 26.5$, which is in fact within the magnitude of the LBG candidate C-4636 from our results with $J_{125} = 26.41$ (see Table \ref{table-phot}).}{For galaxies in GOODS at $J_{125} < 26.5$, i.e. in the range of C-4636 ($J_{125} = 26.41$; see Table \ref{table-phot}), we obtain the highest recovery rate of 43\%.} \edit1{The completeness drops to $\sim10\%$ between $J_{125}=27.5-28.0.$} However, we note that our $5\sigma$ limiting magnitude in this band is deeper reaching up to $J_{125,5\sigma} = 27.46$ (see dashed line in Figure \ref{fig_completeness}). \\

\subsection{Exploring the Environment of \Pisco}\label{overdensity}
\begin{figure}[pt!]
\centering
\includegraphics[width=\linewidth]{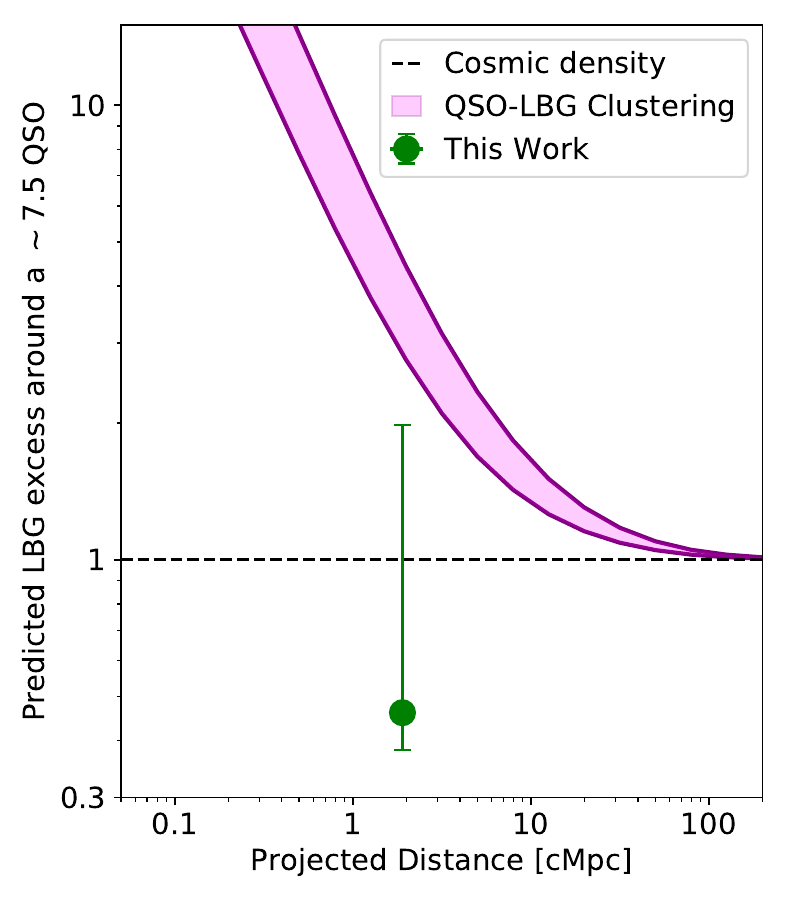}
\caption{Predicted LBG excess as a function of radius around our quasar at $z=7.54$. The pink curve represents the LBG excess taking into account the uncertainty on the determination of the QSO-LBG clustering cross-correlation $r^{QG}_0$ from \cite{garcia-vergara_strong_2017}, assuming no evolution between $z=4$ and $z=7.5$ \edit2{(see \S \ref{overdensity})}. \edit1{Uncertainties due to cosmic variance are not considered.} Accounting for our completeness and the LBG we found at $z=7.69$ with projected distance to the quasar of 223\,pkpc (1.9\,cMpc), we calculate \edit2{an LBG excess of $0.46^{+1.52}_{-0.08}$} (green circle with $1\sigma$ errors from \citealt{gehrels_confidence_1986}), consistent with an average or low-density field. We note that our result is limited by low Poisson statistics.}
\label{qso-lbg_clustering}
\end{figure}

\edit2{In the context of clustering, the probability of finding an excess of LBGs around a quasar is determined by the two--point correlation function, represented as $1+\xi_{QG}(r)$. Here the quasar and LBG (QSO-LBG) cross-correlation is expressed in a power-law form $\xi_{QG}(r)=\left(r/r^{QG}_0\right)^{-\gamma}$, where $r^{QG}_0$ signifies the cross-correlation length and $\gamma$ denotes the slope of the function.}
The highest redshift at which the QSO-LBG clustering has been studied is $z\sim4$ \citep{garcia-vergara_strong_2017}, resulting in $r^{QG}_0=8.83h^{-1}$\,cMpc with a fixed slope $\gamma=2.0$. \edit1{We adopt these measurements and assume} no evolution of the QSO-LBG cross correlation between $z=4$ and $z=7.5$ (830\,Myr). \edit2{The resulting LBG excess as a function of comoving radius around \Pisco\ is presented with the magenta curve in Figure \ref{qso-lbg_clustering}. A quasar field presenting a QSO-LBG excess consistent or above this curve would be considered to reside in a high density region, suggestive of an overdensity. We calculate the excess based on our observation of one QSO-LBG pair relative to the expected number of LBGs in the field. This expected number is calculated from: the number density of LBGs at $z=7.5$ as interpolated from the $z=7$ and $z=8$ rest-UV luminosity functions from \citet{finkelstein_evolution_2015}; our completeness fraction from \S \ref{completeness}; and the comoving cylindrical volume with a radius equivalent to the comoving distance from the quasar to candidate C-4636 and} \edit1{a comoving line-of-sight from the quasar's redshift $\Delta z=\pm0.3$, corresponding to the redshift uncertainties on the C-4636 $z_{phot}$ estimated with \texttt{EAZY} (see \S\ref{z7p5} and Table \ref{table-pz}). Note that in this estimation we do not \edit2{account for candidate} C-\cii\ identified with ALMA in the environment of the quasar, given that this galaxy is not UV-bright.} 

\edit2{Figure \ref{qso-lbg_clustering} shows that at the comoving radius to C-4636 corresponding to 1.9 cMpc,} we observe an LBG excess of $0.46^{+1.52}_{-0.08}$ \edit2{(green circle with error bars)}, indicating that this quasar field is \edit2{consistent with cosmic density (black dashed line where LBG excess = 1), or slightly underdense. This field is incompatible} with an overdensity of UV-bright galaxies as our result is at least 1.4 times below the clustering expectations (magenta curve) around $z=4$ quasars from \citet{garcia-vergara_strong_2017}. Note that this measurement is limited by low number statistics. Therefore, a larger area coverage or optimal set of filters to improve the completeness of LBGs, and spectroscopic follow-ups, would be necessary to fully characterize the environment of this quasar.

JWST observations from program GTO 1219 (PI: Luetzgendorf) aim at confirming the LBG candidate C-4636 at $z\sim7.5$ using NIRSpec MSA spectroscopy. The observations cover 0.7$\mu$m to 3.1$\mu$m with the G140H/F070LP and G235H/F170LP grating and filter combination. \edit1{This setup offers a high resolution power of $\sim 1,000$ and $\sim 2,700$, respectively, \replaced{making the observations sensitive to UV metal emission lines}{enabling sensitivity to UV metal emission lines} such as \civ\,$\lambda$1549, \ciiiS\,$\lambda\lambda$1907,1909 and \mgii\,$\lambda$2798. These lines would characterize the ionization and chemical enrichment of the galaxy \citep[e.g.][]{hutchison_near-infrared_2019}. The Lyman-break at $z\sim 7.5$ from the galaxy would potentially be observed providing additional confirmation of the candidate.}

\added{Recent studies using \jwst\ NIRCam/WFSS spectra have demonstrated overdensities around quasars at slightly lower redshifts. \cite{wang_spectroscopic_2023} found 10 \oiii\ emitting galaxies in the environment of the quasar J0305--3150 at $z=6.6$, probing an overdensity of galaxies in this field. Furthermore, \cite{kashino_eiger_2023} compiled a comprehensive catalog of \oiii\ emitting galaxies at $5.3 < z < 7.0$ in the field of the quasar J0100+2802 at $z = 6.327$. Among the 117 \oiii\ emitters, 24 were associated with the quasar environment, revealing a clear overdensity of galaxies. These findings therefore demonstrate the efficacy of investigating quasar environments by observing strong UV-rest emission lines of galaxies. Applying a similar strategy to the $z=7.54$ quasar \Pisco, the use of NIRCam/WFSS with the F430M filter would be suitable for identifying such \oiii\ emitting galaxies.}

Even though simulations show that massive quasars such as \Pisco\ are good indicators of galaxy overdensities, the opposite has also been observed \citep[e.g.][]{banados_galaxy_2013, simpson_no_2014, mazzucchelli_no_2017}. There is no evidence for an evolutionary trend of overdensities with redshift as for example \cite{mignoli_web_2020} found both LBGs and LAEs in the environment of a quasar at $z=6.31$. On the contrary, \citealt{goto_no_2017} did not find any LAEs around a $z=6.4$ quasar and points out to the possibility that the quasar formation drains out the available matter within the $\sim$ 1 pMpc. There are indeed many physical processes at play in the formation of a quasar and its environment. A possible method to suppress or delay star/galaxy formation within a few pMpc from the quasar is its UV radiation \edit2{\citep[e.g.][]{ota_large-scale_2018,costa_hidden_2019,lambert_lack_2024}.}

In a different scenario, supernova-driven galactic winds could simply sparse the galaxies further away from the quasar and thus reduce the number density observed (e.g. by a factor of up to 3.7 in the \hst/ACS area; \citealt{costa_environment_2014}). Finally, there is also the possibility of the environment being fully dominated by dust-obscured galaxies, and no LBGs or LAEs can be found with traditional photometric techniques. In order to probe this \added{scenario} for the \Pisco\ field, further ALMA observations covering a larger area could unveil this population of galaxy candidates. One can further explore their chemical properties by, e.g., rest-frame optical observations with JWST \cite[e.g][]{decarli_rapidly_2017, garcia-vergara_alma_2022}.

\subsection{Galaxy-Absorber Association at $z\sim 6.8$}
Analysis of the $z=6.84$ absorber detected in the spectrum of \Pisco\ by \citet{simcoe_interstellar_2020} suggests that this system may be classified as a Damped Lyman Alpha (DLA) system, with a fiducial column density of $N_\mathrm{HI}=10^{20.6}$\,cm$^{-2}$ \citep[][]{simcoe_interstellar_2020}. Galaxies originating such absorbers at $z\sim4$ are typically located at impact parameters of $\lesssim50$ pkpc \citep[e.g.,][]{neeleman_c_2017,neeleman_c_2019}. However, recent studies based on $z<2$ MgII ($\lambda\lambda2796,2803\AA$) absorbers, showed that group environment may give rise to stronger and more widespread absorption systems \edit1{within a projected distance of $\lesssim480$ pkpc} \citep[e.g.,][]{nielsen_magiicat_2018, fossati_muse_2019, dutta_muse_2020}. If this behavior holds at high redshift \citep[see][ for implication on the end of the reionization on absorption systems]{doughty_environments_2023}, this can explain the  relatively large impact parameters observed for our two candidates C-4966 and C-5764 (290\,pkpc, 476\,pkpc, assuming $z=6.9$, respectively). The two $z\sim6.9$ galaxy candidates need to be spectroscopically confirmed to establish the physical link with the metal absorption system detected at $z\sim6.8$ by \citet{simcoe_interstellar_2020}.

\section{Summary} \label{summary}
We present the results of a search for Lyman-break galaxy candidates (LBGs) in the environment of the $z=7.54$ quasar \Pisco. We used \hst+\spitzer/IRAC observations designed to look for LBGs in the $\sim1$ proper-Mpc$^2$ environment of the quasar. Here, we present newly obtained deep HST ACS/WFC $i_{814}$ and WFC3 $Y_{105}$ and $J_{125}$ bands. We use the \hst\ observations to select LBG candidates with photometric redshift $z\sim7.5$. Shallower \spitzer/IRAC 3.6 $\mu$m and 4.5 $\mu$m observations are utilized to constrain the high-redshift solution of the galaxies selected.

The final catalog results in the recovery of the quasar and one LBG \added{at $z=7.69$,} with magnitude $J_{125} = 26.4$ and at a projected distance of only 223 pkpc from the quasar. An additional candidate \replaced{associated with the environment of \Pisco\ that had been previously identified with}{previously identified in the environment of \Pisco\ using} ALMA band 6 observations and with $z_{\mathrm{[C\,{\sc II}]}}=7.5341\pm0.0009$ \citep{venemans_kiloparsec-scale_2020} is not detected in any of the five bands used in this work. \replaced{Nevertheless, this is still a potential dust-obscured star-forming galaxy candidate}{This is a potential dust-obscured star-forming galaxy candidate} at $z=7.5$ just 27 pkpc in projection from the quasar.

Galaxy candidates at lower photometric redshifts $z=6.91$ and $z=6.89$ are identified in the data set and, interestingly, are at a redshift that is consistent with a $z=6.84$ absorber in the line of sight previously identified in the quasar spectrum in \citet{simcoe_interstellar_2020}.

The completeness of galaxy candidates found at $z\sim7.5$ in our survey compared to blank fields from GOODS \citep{finkelstein_evolution_2015}, proves to be low even at the brightest magnitudes $J_{125} < 26.5$ ($\sim40\%$). This low completeness can be explained by the fact that a $z\sim7.5$ LBG begins to drop out halfway through the $Y_{105}$, \replaced{biasing results to candidates with red}{leading to biased results favoring candidates with redder} $Y-J$ colors. \added{Taking into account this caveat, we investigate the Quasar-LBG clustering in this field following the studies at $z\sim4$ in \citet{garcia-vergara_strong_2017} and assuming no evolution for clustering. We find that this quasar field is not consistent with an overdensity of LBGs, but \edit2{instead with cosmic density or even an underdense region, noting that this result is heavily influenced by the limitations imposed by Poisson statistics given the sample of only one LBG candidate}. This outcome is puzzling considering the recent findings of overdense quasar environments at $z=6.3$ and $6.6$ in \citet{kashino_eiger_2023, wang_spectroscopic_2023}, respectively. The limitations show that spectroscopy might be crucial as these studies looked for galaxies emitting \oiii\ rather than relied on the \lya\ signature.}

\replaced{The findings from this work are paving the way for follow-up studies that will confirm one of the earliest overdensities in the universe. Spectroscopic confirmation of the LBG candidate at $z\sim7.5$ will be secured with upcoming \jwst\ NIRSpec observations from program GTO 1219.}{The quasar \Pisco\ is one of the most extreme objects in the universe and there are truly several strategies to further explore its environment. First, our} work demonstrates that it is expected to find more \added{LBG} galaxy candidates using further \hst\ or \jwst\ \added{with a more complete set of filters in the }near-infrared. Alternatively, ALMA mosaic observations covering the quasar field \replaced{would be able to unveil the potential population of dust-obscured galaxies.}{could reveal a potential population of dust-obscured galaxies.} \added{Additionally, we could rely on the power of \jwst\ spectra to find galaxies in the field of \Pisco\ by looking for their \oiii\ emission. Finally, expanding the search of galaxies to a wider area of up to 10 comoving-Mpc could probe necessary to \replaced{test}{thoroughly investigate} the environment of this $z=7.54$ quasar \citep{overzier_realm_2016, chiang_ancient_2013}.}\\


\acknowledgements
We thank the anonymous referee for the valuable comments and inquiries that enriched this manuscript. Acknowledgment is extended to the help desk at Space Telescope Science Institute for their support in data reduction. S.R.R. acknowledges financial support from the International Max Planck Research School for Astronomy and Cosmic Physics at the University of Heidelberg (IMPRS--HD). S.R.R and C.M. acknowledge the Science Support Discretionary Fund at the European Southern Observatory, Vitacura, Chile. E.P.F. is supported by the international Gemini Observatory, a program of NSF’s NOIRLab, which is managed by the Association of Universities for Research in Astronomy (AURA) under a cooperative agreement with the National Science Foundation, on behalf of the Gemini partnership of Argentina, Brazil, Canada, Chile, the Republic of Korea, and the United States of America.  We would like to thank Nina Hatch for sharing her expertise in protoclusters and overdensities of galaxies at high redshift. This paper makes use of \hst\ data from program cy15/GO 1165 that can be found in MAST: \dataset[10.17909/q77c-8j50]{http://dx.doi.org/10.17909/q77c-8j50}, \textit{Spitzer}/IRAC cy16/13201 that can be found in \dataset[10.5281/zenodo.8389635]{https://zenodo.org/records/8389635}, and ALMA$\#$2017.1.00396.S\\

\facilities{\hst, \emph{Spitzer/IRAC}, ALMA}
\software{Astropy \citep{the_astropy_collaboration_astropy_2013},
          DrizzlePac \citep{hack_drizzlepac_2013},
          EAZY \citep{brammer_eazy_2008},
          GOLFIR \citep{brammer_gbrammergolfir_2022}
          JS9-4L \url{ https://waps.cfa.harvard.edu/eduportal/js9/software.php},
          Matplotlib \citep{hunter_matplotlib_2007},
          Numpy \citep{harris_array_2020},
          PyPHER \citep{boucaud_convolution_2016},
          SciPy \citep{virtanen_scipy_2020},
          Source Extractor \citep{bertin_sextractor_1996}
          }

\bibliographystyle{yahapj}
\bibliography{references.bib}          
\end{document}

%% file: definitions.tex
\definecolor{red}{rgb}{1,0,0}
\definecolor{orange}{RGB}{204, 85, 0}
\definecolor{blue}{HTML}{4169e1}
\definecolor{ltred}{RGB}{245,167,162}
\definecolor{ltblue}{RGB}{206,211,242}

\newcommand{\lya}{Ly$\alpha$}

\newcommand{\oiii}{[\ion{O}{3}]}

\newcommand{\civ}{C\,{\sc iv}}

\newcommand{\cii}{[C\,{\sc ii}]}

\newcommand{\ciiiS}{C\,{\sc iii}]}
\newcommand{\mgii}{Mg\,{\sc ii}}

\newcommand{\Msun}{$M_\odot$}

\newcommand{\hst}{\textit{HST}}
\newcommand{\jwst}{\textit{JWST}}
\newcommand{\spitzer}{\textit{Spitzer}}

\newcommand{\Pisco}{ULAS J1342+0928}